\documentclass{article}
\usepackage{newtxtext,newtxmath}
\usepackage{amsmath}
\usepackage{graphicx}
\usepackage{multicol}
\usepackage{titlesec}
\usepackage{enumitem}
\usepackage{bm}
\usepackage{array}
\usepackage[a4paper, margin=2cm]{geometry}
\usepackage{longtable}
\usepackage{graphicx} % 用于包含图像
\usepackage{float}    % 提供 H 选项
\usepackage{caption} 
\usepackage{ragged2e}
\usepackage{fancyhdr}
\emergencystretch=\maxdimen
\renewcommand{\normalsize}{\fontsize{10.5}{14}\selectfont}
\titlespacing*{\section}{0pt}{2ex}{0.5ex}
\titlespacing*{\subsection}{0pt}{1ex}{0ex}
\titleformat{\section} % Command
  {\normalfont\normalsize\centering} % Format: 粗体，居中
  {\Roman{section}.} % Label: 使用罗马数字
  {0em} % Sep: 标题和标题名之间的间隔
  {} % Before-code

\renewcommand{\thesubsection}{\arabic{subsection}}
\makeatletter
\@addtoreset{subsection}{section}
\makeatother
\titleformat{\subsection} % Command
  [hang] % Shape
  {\normalfont\itshape\fontsize{10.5}{12}\selectfont} % Format: Times New Roman，10.5pt 字号，斜体，不加粗
  {\thesubsection} % Label: 使用数字 1, 2, 3 等
  {0em} % Sep: 标题和标题名之间的间隔
  {} % Before-code
\title{\textbf{An Improved Reversible Data Hiding Algorithm Based on Reconstructed Mapping for PVO-k}}
\author{Yusen Zhang\textsuperscript{1}, Haoyun Xu\textsuperscript{2}, Jingwen Li\textsuperscript{2*} \\
\textsuperscript{1}\textsuperscript{2}University of Shanghai for Science and Technology, China \\
\text{cbs\_edu@outlook.com}}
\date{}

\begin{document}

\maketitle
\thispagestyle{empty}

\noindent\textbf{\textit{Abstract—}Reversible Data Hiding (RDH) is a practical and efficient technique for information encryption. Among its methods, the Pixel-Value Ordering (PVO) algorithm 
and its variants primarily modify prediction errors to embed information. However, both the classic PVO and its improved versions such as IPVO and PVO-k, share a common limitation: 
their maximum data embedding capacity for a given grayscale image is relatively low. This poses a challenge when large amounts of data need to be embedded into an image. In response to these issues,
 this paper proposes an improved design targeting the PVO-k algorithm. We have reconstructed the mapping scheme of the PVO-k algorithm to maximize the number of pixels that can embed encrypted information. 
 Experimental validations show that our proposed scheme significantly surpasses previous algorithms in terms of the maximum data embedding capacity. For instance, when embedding information into a grayscale image of an airplane, 
  our method's capacity exceeds that of PVO-k by 11,207 bits, PVO by 8,004 bits, and IPVO by 4,562 bits. The results demonstrate that our algorithm holds substantial advantages over existing methods and introduces innovative mapping ideas, laying a foundation for future research in reversible data hiding in images.}

\noindent\textbf{\textit{Keywords—}Reversible Image Data Hiding; Pixel Value Ordering; Mapping Reconstruction; High-capacity Embedding}
  
\begin{multicols}{2}
\section{INTRODUCTION}
\hspace{1.2em}
\justifying
Embedding encrypted information within images is a common technique in the field of information security. Traditional information embedding techniques can cause permanent damage to the carrier image. However, in many practical applications, it is necessary not only to embed encrypted information into the carrier image but also to ensure that the carrier image can be perfectly restored after the encrypted information is extracted. This requirement has given rise to the technology of reversible image steganography, which allows information to be embedded into an image and ensures that the image can be totally recovered after the embedded information is retrieved.

Under the concept of reversible information hiding, researchers have developed the Pixel Value Ordering (PVO) algorithm [1]. This algorithm constructs a one-to-one reversible mapping based on the relationship between the maximum and the second maximum grayscale values of pixels within each image block, thus achieving reversible information embedding. While the PVO algorithm is fundamentally robust, it has some shortcomings [2]–[7]. Consequently, researchers introduced the Improved PVO (IPVO) algorithm [8], which enhances the mapping method of the original PVO algorithm and increases the maximum information embedding capacity. Furthermore, the PVO-k algorithm was designed to embed encrypted information into the image while minimizing the disruption to the carrier image, thereby reducing the distortion of the encrypted image [9]–[15]. Although the PVO-k algorithm reduces image distortion, it also significantly decreases the maximum information embedding capacity, leading to a decline in overall performance[16]–[23]. Therefore, based on the design principles of the PVO-k algorithm, we have improved it by developing a new mapping pattern that allows as many pixels as possible within an image block to be used for embedding encrypted information. Experimental results, using a grayscale image of an 
airplane as an example, demonstrate that our proposed scheme can achieve a maximum embedding capacity exceeding that of PVO-k by 11,207 bits, PVO by 8,004 bits, and IPVO by 4,562 bits. These results show that our algorithm has substantial advantages over the existing methods.

This article will be divided into five parts. The first part is the introduction, which provides an overview of the article. The second part introduces related work, where we will briefly explain the basic principles of the PVO, IPVO, and PVO-k algorithms, laying the groundwork for the proposal of our scheme. The third part will introduce our proposed scheme, detailing the design ideas, working principles, and algorithmic processes. The fourth part will discuss the experimental performance of our algorithm, where we will compare our algorithm's performance advantages over PVO, IPVO, and PVO-k using standard 512x512 grayscale images such as Lena, Airplane, Lake, Barbara, Pepper, and Baboon to test the performance of the algorithm under different image texture conditions. The fifth part is our conclusion and expectations, which discusses the directions for further research and potential challenges we may face.

\section{RELATED WORK}
\subsection{.PVO Algorithm}
\hspace{1.2em}
The Pixel Value Ordering (PVO) algorithm embeds information into images by initially segmenting a carrier image into uniformly sized blocks. Each image block, denoted as $I$ , is defined as follows:
\begin{gather}
    I = \{ P_i \vert i=1,2...,n\} 
\end{gather}

where $P_i$ represents the pixels that make up the block, and \textbf{n} is the number of pixels in that block. The pixels within a block are first sorted in descending order based on their grayscale values. We define a mapping $\gamma(n)$ , where $P_{\gamma(n)}$ represents the pixel with the highest grayscale value in the block, and $P_{\gamma(n-1)}$ represents the pixel with the second highest grayscale value, and so on. After sorting the pixels in descending order, we obtain the following sequence of pixel values:
\begin{gather}
    P_{\gamma(n)}\geq p_{\gamma(n-1)}\geq...\geq P_{\gamma(2)}\geq p_{\gamma(1)}
\end{gather}

Next, we extract the pixel with the highest grayscale value $P_{\gamma(n)}$,
and the pixel with the second highest grayscale value, $P_{\gamma(n-1)}$from the image block.
We then calculate the predictive error $e_{max}$ for this image block. The specific definition of $e_{max}$ is as follows:
\begin{gather}
    e_{max} = p_{\gamma(n)} - P_{\gamma(n-1)}
\end{gather}

After calculating the specific value of $e_{max}$,
within the image block, we modify the grayscale value of the pixel with the highest grayscale $P_{\gamma(n)}$,
according to the value of $e_{max}$.The modification is performed as it shown in Equation (4) to embed the information:
\begin{gather} 
\widetilde{P}_{\gamma(n)} = \begin{cases}
    P_{\gamma(n)}, & e_{\text{max}} = 0 \\
    P_{\gamma(n)} + b, & e_{\text{max}} = 1 \\
    P_{\gamma(n)} + 1, & e_{\text{max}} > 1
\end{cases}
\end{gather}

In this process, $b$ represents the binary bit of information to be embedded. By executing the above operation on each image block, we can embed a string of binary encrypted information into the carrier image.

Next, we will focus on the decryption of the encrypted image and the restoration of the carrier image. Observing Equation (4), it is evident that after modifying the pixel with the highest grayscale value, this pixel remains the one with the highest grayscale value in the entire image block. This ensures the stability of the grayscale value ordering before and after modification, providing a prerequisite for constructing a reversible map.

The decryption of the image is conducted as follows: After obtaining the encrypted image, we continue to divide the image into blocks according to the encryption rules. Then, for each image block, we sort the pixels inside according to the grayscale values in descending order, obtaining the sorted sequence as it shown in follows:
\begin{gather}
    \widetilde{P}_{\gamma(n)} \geq p_{\gamma(n-1)}\geq...\geq P_{\gamma(2)}\geq p_{\gamma(1)}
\end{gather}

Next, we calculate the new predictive error.
\begin{gather}
    \widetilde{e}_{max} = \widetilde{P}_{\gamma(n)}-P_{\gamma(n-1)}
\end{gather}

After calculating the new predictive error $\widetilde{e}_{max}$, it is easy to observe the following:

\begin{itemize}[leftmargin=*]
    \setlength{\itemsep}{1pt}
\setlength{\parsep}{1pt}
\setlength{\parskip}{1pt}
\item If $\widetilde{e}_{max} = 0$, it indicates that $e_{max}=0$,meaning no encrypted information was embedded.
\item If $\widetilde{e}_{max} = 1$, signifies that $e_{max}=1$, and the embedded information $b=0$
\item If $\widetilde{e}_{max} = 2$, this implies that $e_{max}=1$ and the embedded information $b=1$
\item If $\widetilde{e}_{max} > 2$, it indicates that $e_{max}=\widetilde{e}_{max}-1$, with no encrypted information embedded.
\end{itemize}

This way, we can smoothly extract encrypted information from the image blocks, and restoring the carrier image is also straightforward. We have already deduced the
$e_{max}$ values for the image blocks. According to Equation (3), we can deduce that:
\begin{gather}
    P_{\gamma(n)}=P_{\gamma(n-1)} +e_{max}
\end{gather}

And based on Equation (7), we can restore the recently modified $\widetilde{P}_{\gamma(n)}$, After restoring $\widetilde{P}_{\gamma(n)}$ for each image block, the entire image can then be totally restored.\\

\subsection{.IPVO Algorithm}
\hspace{1.2em}
The Pixel Value Ordering (PVO) algorithm is generally well-developed but exhibits some shortcomings in performance in certain aspects. The Improved PVO (IPVO) algorithm addresses these deficiencies, offering an enhancement particularly in the maximum capacity for encrypted information embedding compared to the original PVO algorithm. The implementation process of the IPVO algorithm is largely similar to that of PVO. It follows the initial steps of sorting the pixels within an image block by their grayscale values, resulting in a descending order sequence as shown in Equation (2), A further explanation is required for the mapping $\gamma(n)$,
which represents the position of the pixel with the highest value in the sequence before sorting.

The method of calculating the predictive error in the IPVO algorithm differs from that in the PVO algorithm. In the IPVO algorithm, the predictive error is defined as shown in Equation (8)
\begin{gather}
    e_{max} = P_u - P_\nu 
\end{gather}
where
\begin{center}
    $u = min(\gamma(n),\gamma(n-1))$

$\nu = max(\gamma(n),\gamma(n-1))$
\end{center}

Next, we modify $e_{max}$ according to the mapping relationship constructed in Equation (9).
\begin{gather} 
    \widetilde{e}_{\text{max}} = \begin{cases}
        e_{\text{max}} - 1, & e_{\text{max}} < 0 \\
        e_{\text{max}} - b, & e_{\text{max}} = 0 \\
        e_{\text{max}} + b, & e_{\text{max}} = 1 \\
        e_{\text{max}} + 1, & e_{\text{max}} > 1
    \end{cases}
\end{gather}

Essentially, modifying $e_{max}$ sill involves adjusting $P_{\gamma(n)}$,
but unlike the PVO algorithm, the IPVO algorithm incorporates the indices of the original pixel sequence, adding new conditions for auxiliary storage. The PVO algorithm can only embed encrypted bits into an image block when it $\widetilde{e}_{max}$
is calculated as 0. In contrast, the IPVO algorithm allows for the embedding of encrypted bits when $e_{max}$ is calculated as 0 or 1, thereby enhancing the volume of embedded information.

Decryption of images encrypted using the IPVO algorithm is conducted as follows: first, the pixels in the encrypted image block are sorted in descending order of grayscale values. Then, based on the calculated $\widetilde{e}_{max}$
and using Equation (9), we can infer the value of $e_{max}$, as well as the embedded information $b$, as shown in Equations (10) and (11)
\begin{gather} 
    e_{\text{max}} = \begin{cases}
        0, & \widetilde{e}_{\text{max}} = 0 , -1 \\
        1, & \widetilde{e}_{\text{max}} = 1, 2 \\
        \widetilde{e}_{\text{max}} + 1, & \widetilde{e}_{\text{max}} < -2 \\
        \widetilde{e}_{\text{max}} - 1, & \widetilde{e}_{\text{max}} > 2
    \end{cases}
\end{gather}

\begin{gather} 
    b = \begin{cases}
        0, & \widetilde{e}_{\text{max}} = 0, 1 \\
        1, & \widetilde{e}_{\text{max}} = -1, 2
    \end{cases}
\end{gather}

The method for restoring images in the IPVO algorithm is also relatively straightforward. In practice, the IPVO algorithm still modifies the pixel with the highest grayscale value in the image block,
$\widetilde{P}_{\gamma(n)}$.  As indicated by Equation (8)

\noindent If $\gamma(n) > \gamma(n-1)$, then $e_{max} = P_{\gamma(n-1)}-P_{\gamma(n)}$ is less than 0.

\noindent If $\gamma(n) < \gamma(n-1)$, then $e_{max} = P_{\gamma(n)}-P_{\gamma(n-1)}$ is greater than 0.

Based on this reasoning, the calculation method for $P_{\gamma(n)}$ can be derived as shown in Equation (12)
\begin{gather}
P_{\gamma(n)} = \begin{cases}
    P_{\gamma(n-1)} - e_{\text{max}}, & e_{\text{max}} < 0 \\
    P_{\gamma(n-1)} + e_{\text{max}}, & e_{\text{max}} > 0
\end{cases}
\end{gather}
\end{multicols}
\subsection{.PVO-k Algorithm}
\hspace{1.2em}
From the introduction of the PVO algorithm, it is evident that when the grayscale values of two pixels,
$P_{\gamma(n)}$ and $P_{\gamma(n-1)}$,
within an image block are equal, it is impossible to embed encrypted information into that block, resulting in a certain waste of space. To address this issue, researchers have developed the PVO-k algorithm. The PVO-k algorithm focuses on scenarios where the first
$k$ elements in the pixel grayscale value sequence are equal like (13).
\begin{gather}
    P_{\gamma(n)} = P_{\gamma(n-1)}=...=P_{\gamma(n-k+1)}\geq P_{\gamma(2)}\geq P_{\gamma(1)}
\end{gather}

\begin{figure}[H]
    \centering
    \setlength{\abovecaptionskip}{0.5cm}
    \setlength{\belowcaptionskip}{-0.1cm} 
    \includegraphics[width=1\textwidth]{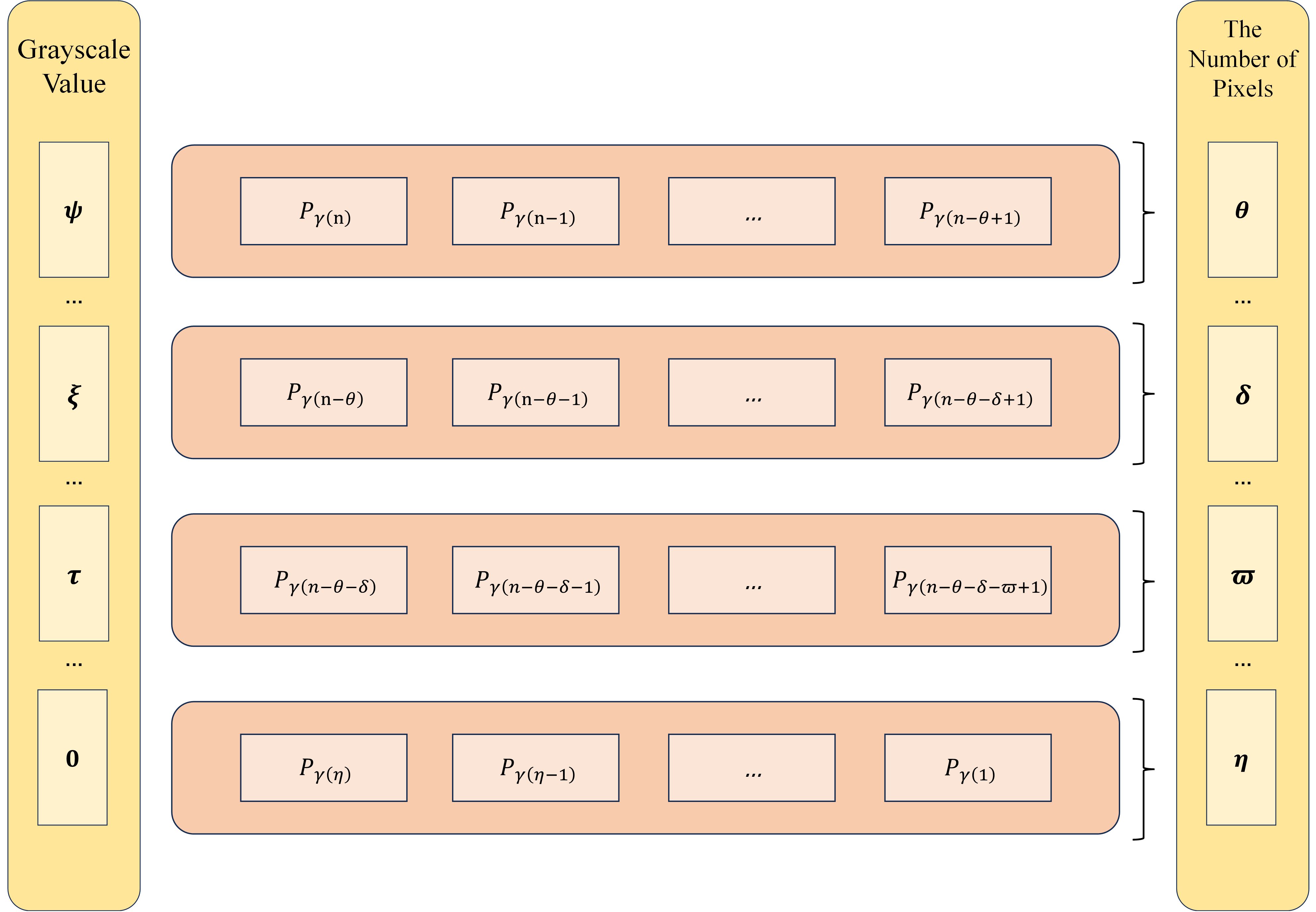}
    \caption{Arrangement of pixels in grayscale images}
    \end{figure} 

\begin{multicols}{2}
    Fig.1 displays the arrangement of pixels in a grayscale image, where $\psi,\xi,\tau,o$ represent the sizes of the grayscale values of these pixels, 
    satisfying the relationships $\psi >\xi >\tau >o$, and $\theta ,\delta ,\varpi ,\eta$ denote the quantities of pixels whose grayscale values correspond to
    $\psi,\xi,\tau,o$, respectively. In the PVO-k algorithm, the predictive error $e_{max}$ is defined by Equation (14)
\begin{gather}
    e_{max} = P_{\gamma(n)} - P_{\gamma(n-k)}
\end{gather}

If the calculated $e_{max}$ for an image block is 1, we proceed with embedding encrypted information. It is evident that the PVO-k algorithm performs exceptionally well when embedding information into smooth images.

During the embedding process, we modify the first $k$ pixels that have the highest and equal grayscale values,
$P_{\gamma(i)}$,  to facilitate the information embedding, where $i\in [n-k+1,n]$,
The method of modification is shown in Equation (15)
\begin{gather}
    \widetilde{P}_{\gamma(n)} = \begin{cases}
        P_{\gamma(i)} + b, & e_{max} = 1 \\
        P_{\gamma(i)} + 1, & e_{max} > 1
    \end{cases}
    \end{gather}

The method of information extraction and restoration of the carrier image in the PVO-k algorithm is largely similar to the methods described in the previously introduced algorithms; therefore, this paper will not elaborate further on that.\\

\section{PROMOTED PVO-K ALGORITHEM}
\hspace{1.2em}
In the previous section, we briefly discussed the prior research on algorithms related to PVO. In this section, we will thoroughly introduce the core theory and implementation methods of our proposed Promoted PVO-k (PPVO-k) algorithm.

Whether it is the PVO, IPVO, or PVO-k algorithms, while they are relatively well-developed, their performance in terms of maximum information embedding capacity still has room for improvement. Taking the PVO-k algorithm as an example, although it utilizes image blocks where
$P_{\gamma(n)} = P_{\gamma(n-k)}=...=P_{\gamma(n-k+1)}$ its use of these $k$
equal grayscale pixels from $P_{\gamma(n)}$ to $P_{\gamma(n-k+1)}$ is not fully efficient.The PPVO-k algorithm, based on this issue, improves upon the PVO-k algorithm. The PPVO-k algorithm makes it possible to embed encrypted information into 
all $k$ pixels within a block, $P_{\gamma(n)}, P_{\gamma(n-k)},...,P_{\gamma(n-k+1)}$, significantly increasing the information embedding capacity of a single image block.

\subsection{.The Information Encryption Process of the PPVO-k Algorithm}
\hspace{1.2em}
First, we define three distinct types of encryption sequences for a set of binary bits that are to be embedded into an image block
\begin{gather}
    S = \{b_i \vert  i =1,2,...,m \}
\end{gather}

If the sequence consists entirely of zeros, we denote it as S0. If the sequence consists entirely of ones, we denote it as S1. If the sequence is a mix of zeros and ones, we denote it as MS.

As shown in Fig.1, we refer to the $\theta $  pixels with the highest identical grayscale values as first-order pixels, denoting their grayscale value as $O1$.
The next highest $\delta $ identical pixels are called second-order pixels, with their grayscale value noted as $O2$,
and so on. In the PPVO-k algorithm, the predictive error $e_{max}$ is defined as follows
\begin{gather}
    e_{max} = O1-O2
\end{gather}

The PPVO-k algorithm stipulates that if the predictive error $e_{max}$ of an image block is calculated to be 1, we will proceed to embed information into that block. Next, we will discuss the various scenarios for embedding encrypted information into image blocks.\\

\noindent\textit{A. Cases where the embedded information is either an all-zero sequence or an all-one sequence:}
\begin{itemize}[leftmargin=*]
    \setlength{\itemsep}{1pt}
\setlength{\parsep}{1pt}
\setlength{\parskip}{1pt}
    \item If the information to be embedded in an image block is an all-zero sequence, we will increase the grayscale values of all first-order pixels in that block by 1. After this modification, the calculated $\widetilde{e}_{max}$ for the block becomes 2.
    \item If the information to be embedded is an all-one sequence, we will increase the grayscale values of all first-order pixels in that sequence by 2. After modification $\widetilde{e}_{max}$ becomes 3.
\end{itemize}
\textit{B. Cases where the embedded information is a mixed sequence of 0 and 1}

If the information to be embedded in the block consists of a mixed sequence of 0 and 1, we first extract all first-order pixels $P_{\gamma(n)}, P_{\gamma(n-k)},..., P_{\gamma(n-k+1)}$.
Next, we sort these pixels by their indices in ascending order, resulting in the sequence $P_{\theta (n)},P_{\theta (n-1)},...,P_{\theta (n-\theta+1)}$,
where $\theta(n)<\theta(n-1)<...<\theta(n-\theta+1)$. Using the sorted sequence, we embed the encrypted information into each pixel
$P_{\theta(i)}$, $i\in [n-\theta+1,n]$ as shown in Fig 2. After embedding the information, we can find that $\widetilde{e}_{max}$ equal 1.

If the calculated $e_{max}$ for an image block is 0, no modifications are made to that block, keeping $\widetilde{e}_{max} = 0$.
If $e_{max}$ exceeds 2, increase the grayscale values of all first-order pixels in the block by 2. This adjustment results in $\widetilde{e}_{max} = e_{max}+2$\\

\subsection{.Information Extraction and Carrier Image Restoration in the PPVO-k Algorithm}
\hspace{1.5em}
Next, we will describe how to extract information from images encrypted using the PPVO-k algorithm and how to restore the carrier image. After obtaining the encrypted image, we divide the image according to predetermined rules, calculate the predictive error $\widetilde{e}_{max}$
for each image block, and then discuss different scenarios based on the embedding results of $\widetilde{e}_{max}$

\noindent\textbf{A. $\bm{\widetilde{e}_{max} = 0 \quad or \quad \widetilde{e}_{max} >3}$}

\begin{itemize}[leftmargin=*,topsep=-2pt]
    \setlength{\itemsep}{1pt}
\setlength{\parsep}{1pt}
\setlength{\parskip}{1pt}
    \item If $\widetilde{e}_{max} =0 $, it indicates that no encrypted information was embedded in the image block, and $e_{max} = \widetilde{e}_{max}$.The image block has not been altered, so no restoration is necessary.
    \item If $\widetilde{e}_{max}>3$,  it suggests that the image block does not contain any embedded encrypted information, and the original $e_{max} = \widetilde{e}_{max} - 2$, To restore this image block, all first-order pixels should have their grayscale values reduced by 2.
\end{itemize}

\noindent\textbf{B. $\bm{\widetilde{e}_{max} = 2 \quad or \quad \widetilde{e}_{max} = 3}$}

When $\widetilde{e}_{max}=2 $,  it indicates that the image block has embedded $\theta$
bits. To restore, retrieve the first-order pixels $\widetilde{P}_{\gamma(n)},\widetilde{P}_{\gamma(n-1)},...,\widetilde{P}_{n-\theta+1}$
from the block and reduce the grayscale value of each pixel by 1, specifically
\begin{gather}
    P_{\gamma(i)} = \widetilde{P}_(\gamma(i)-1), i\in [n-\theta+1,n]
\end{gather}
\end{multicols}
\begin{figure}[H]
    \centering
    \setlength{\abovecaptionskip}{0.5cm}
    \setlength{\belowcaptionskip}{-0.1cm} 
    \includegraphics[width=1\textwidth]{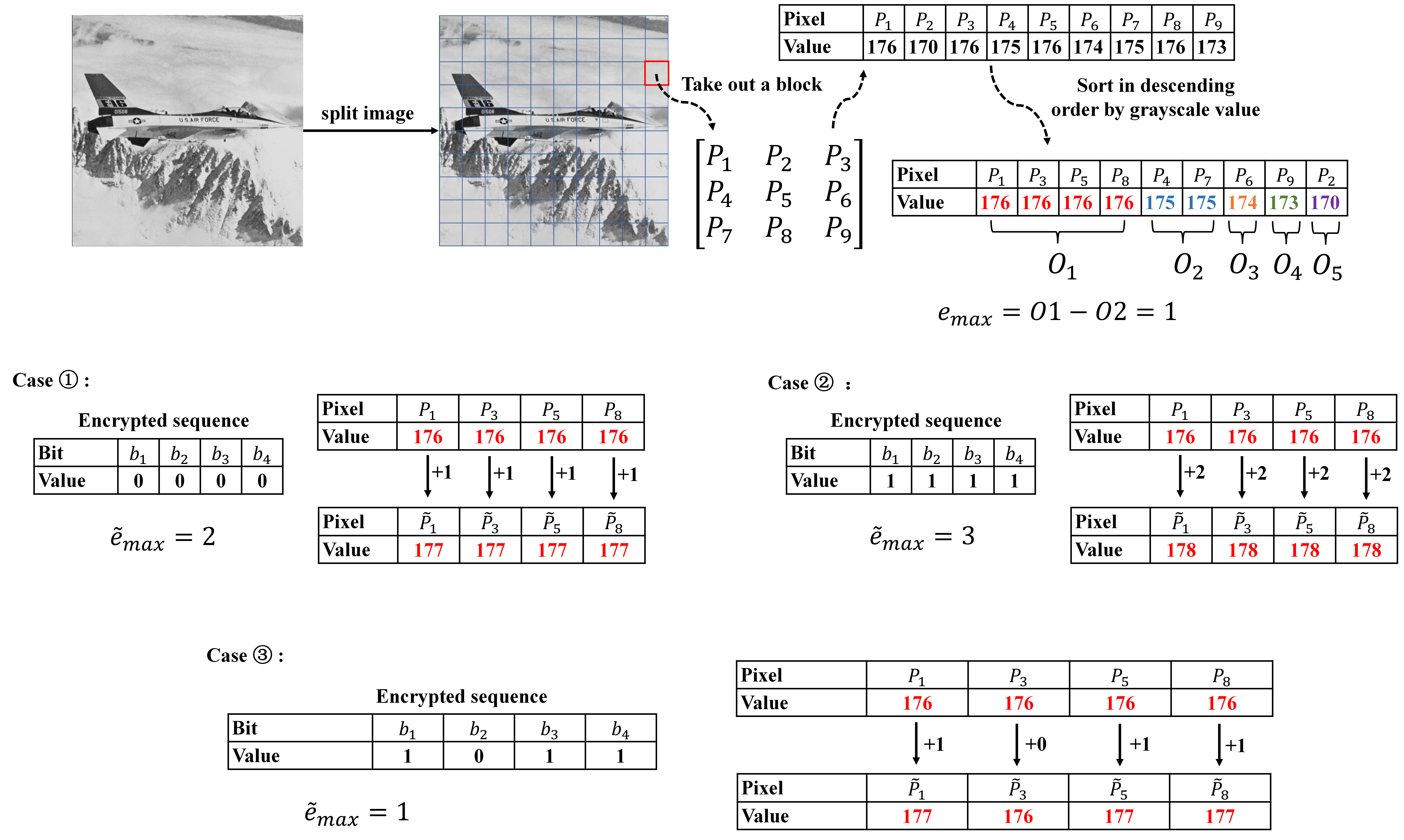}
    \caption{The schematic of the PPVO-k algorithm for encrypted information embedding}
    \end{figure} 

\begin{multicols}{2}
    When $\widetilde{e}_{max} = 3$, it means that the image block has 
    embedded $\theta$ one bits. For restoration, decrease the 
    grayscale values of all first-order pixels in the block by 2. \\
    
\noindent\textbf{C. $\bm{\widetilde{e}_{max} = 1 }$}

Next, we will focus on how to extract information when the embedded encrypted information consists of a mixed sequence of zeros and ones. As explained in the previous section, when $\widetilde{e}_{max} = 1$, it indicates that the image block contains a mixed sequence of zeros and ones.

First, we extract all first-order and second-order 
pixels from the image block, specifically
$\widetilde{P}_{\gamma(n)}, \widetilde{P}_{\gamma(n-1)},...,\widetilde{P}_{\gamma(n-\theta^{'}+1)},...,\widetilde{P}_{\gamma(n-\theta^{'}-\delta^{'}+1)}$,
where $\theta = \theta^{'}+\delta^{'}$. Here, $\theta$ is the number of first-order pixels before modification,
$\theta^{'}$ is the number of first-order pixels after modification,
and $\delta^{'}$is the number of second-order pixels after modification.

Next, we take the extracted $\theta$ pixels and sort them 
in ascending order based on their indices to obtain the 
sequence $\widetilde{P}_{\Theta (n)},\widetilde{P}_{\Theta(n-1)},...,\widetilde{P}_{\Theta(n-\theta+1)}$. We then 
calculate the average grayscale value of this sorted 
sequence.
\begin{gather}
    \bar{P} = \frac{\sum_{i=n-\theta+1}^{n} \widetilde{P}_{\Theta(i)}}{\theta}
\end{gather}

Compare the average grayscale value $\bar{P}$ with each pixel in the sequence $\widetilde{P}_{\Theta (n)},\widetilde{P}_{\Theta(n-1)},...,\widetilde{P}_{\Theta(n-\theta+1)}$.
If $\bar{P}>\widetilde{P}_{\Theta(i)}$, then $\widetilde{P}_{\Theta(i)}$  is classified as a second-order 
pixel.If $\bar{P}<\widetilde{P}_{\Theta(i)}$, then  $\widetilde{P}_{\Theta(i)}$ is classified as a first-order pixel. Based on the results of these comparisons, 
formulas (20) and (21) can be derived.

\begin{gather}
    b_i = \begin{cases}
        0, & if \quad \bar{P} > \widetilde{P}_{\Theta(i)} \\
        1, & if \quad \bar{P} < \widetilde{P}_{\Theta(i)}
    \end{cases}
\end{gather}

\begin{gather}
    P_{\Theta(i)} = \begin{cases}
        \widetilde{P}_{\Theta(i)} , & \bar{P} > \widetilde{P}_{\Theta(i)} \\
        \widetilde{P}_{\Theta(i)} - 1, & \bar{P} < \widetilde{P}_{\Theta(i)}
    \end{cases}
\end{gather}

The above outlines the implementation process of 
the PPVO-k algorithm. Next, we will present a 
comparative analysis of experimental results to 
demonstrate the advantages of the PPVO-k algorithm 
over other traditional algorithms.

\section{EXPERIMENTAL RESULTS}
\hspace{1.2em}
Next, we will conduct performance tests and 
comparisons of our PPVO-k algorithm against the PVO, 
IPVO, and PVO-k algorithms. We have selected six 
typical 512x512 experimental images from the Kodak dataset for information embedding. These images are 
Lena, Airplane, Lake, Barbara, Pepper, and Baboon, as 
shown in Figure 3. Each of these images varies in 
texture and smoothness, representing different types 
commonly encountered in real life.

We sequentially compared the maximum single-instance embedding capacity of the different algorithms 
on each image. By employing a dynamic block size 
selection mechanism, we ultimately chose to use a 
block size of 2x2 for image segmentation. The 
experimental results are presented in Table 1.
\begin{center}
    Table1.
        
    \textit{Maximum Single-Instance Information Embedding 
    Capacity of Different Algorithms on Various Images}
\end{center}
    \begin{center}
        \begin{tabular}{>{\centering\arraybackslash}m{2.2cm} >{\centering\arraybackslash}m{2.2cm} >{\centering\arraybackslash}m{2.2cm}}
            \hline
            Algorithm&Picture & Embedding Capacity (bits) \\ \hline
            PVO & Airplane & 31816 \\ \hline
            IPVO & Airplane & 35258 \\ \hline
            PVO-k & Airplane & 28613 \\ \hline
            PPVO-k & Airplane & 39280 \\ \hline
            PVO & Lena & 20591 \\ \hline
            IPVO & Lena & 21602 \\ \hline
            PVO-k & Lena & 19551 \\ \hline
            PPVO-k & Lena & 23822 \\ \hline
            PVO & Barbara & 15985 \\ \hline
            IPVO & Barbara & 16483 \\ \hline
            PVO-k & Barbara & 16897 \\ \hline
            PPVO-k & Barbara & 20132 \\ \hline
            PVO & Lake & 15004 \\ \hline
        
        \end{tabular}
    \end{center}

    \begin{center}
        \begin{tabular}{>{\centering\arraybackslash}m{2.2cm} >{\centering\arraybackslash}m{2.2cm} >{\centering\arraybackslash}m{2.2cm}}
            \hline
            IPVO & Lake & 15255 \\ \hline
            PVO-k & Lake & 13974 \\ \hline
            PPVO-k & Lake & 16489 \\ \hline
    PVO & Baboon & 8479 \\ \hline
            IPVO & Baboon & 8386 \\ \hline
            PVO-k & Baboon & 7026 \\ \hline
            PPVO-k & Baboon & 7569 \\ \hline
            PVO & Pepper & 17205 \\ \hline
            IPVO & Pepper & 17377 \\ \hline
            PVO-k & Pepper & 16281 \\ \hline
            PPVO-k & Pepper & 18774 \\ \hline
        \end{tabular}
    \end{center}

To more clearly present the experimental results, 
we have visualized the data from Table 1, as shown in 
Figure 4. It is evident from both Table 1 and Figure 4 
that the PPVO-k algorithm outperforms the other 
algorithms on the majority of carrier images. 
Furthermore, the smoother the image, the better the 
performance of the PPVO-k algorithm. For example, in 
the embedding tests on the Airplane image, the 
maximum single-instance embedding capacity of the 
PPVO-k was higher by 11,207 bits compared to PVO-k, 8,004 bits higher than PVO, and 4,562 bits higher 
than IPVO. Conversely, if the carrier image is too 
coarse, the performance of the PPVO-k algorithm is 
adversely affected. For instance, in the embedding tests 
on the Baboon image, the performance of PPVO-k was 
weaker compared to other algorithms. This highlights 
an area for potential improvement that we need to 
consider in the future.
\end{multicols}
\begin{figure}[H]
    \centering
    \setlength{\abovecaptionskip}{0.5cm}
    \setlength{\belowcaptionskip}{-0.1cm} 
    \includegraphics[width=0.8\textwidth]{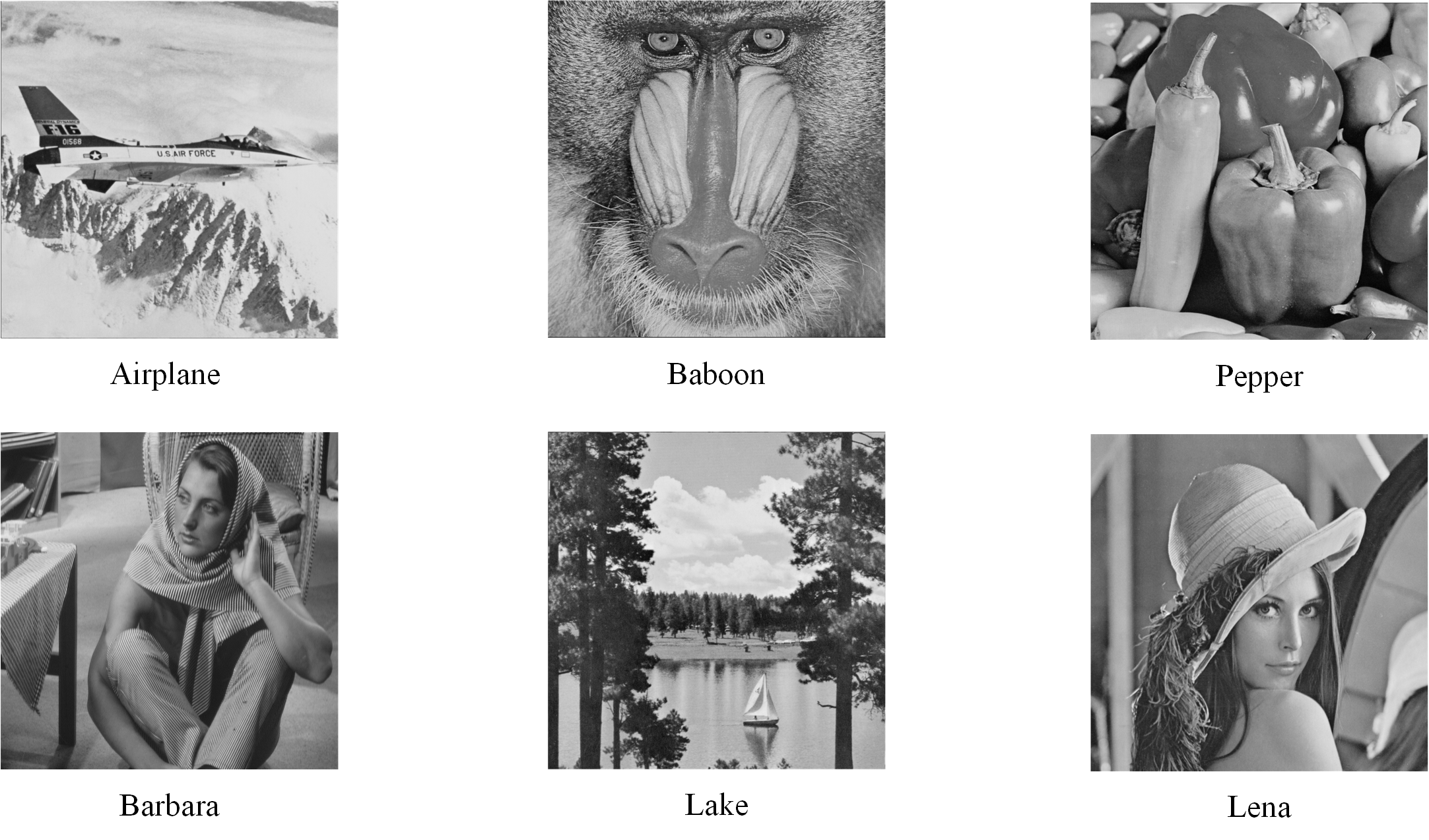}
    \caption{ Six standard 512x512 grayscale images for experiments}
    \end{figure} 
\begin{figure}[H]
        \centering
        \setlength{\abovecaptionskip}{0.5cm}
        \setlength{\belowcaptionskip}{-0.1cm} 
        \includegraphics[width=1\textwidth]{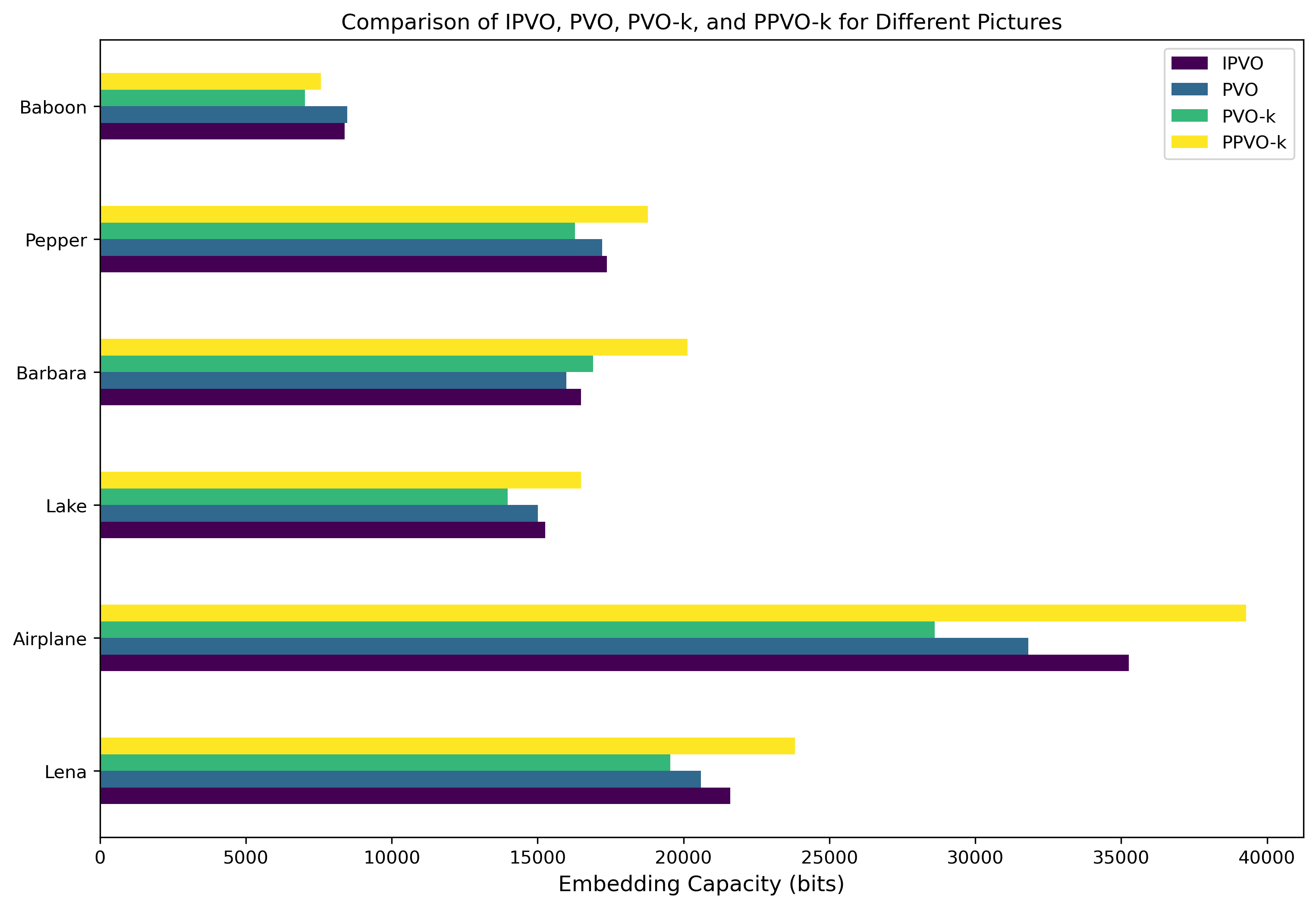}
        \caption{The Comparison of IPVO, PVO, PVO-k, and PPVO-k for Different Pictures}
\end{figure} 

\begin{multicols}{2}
\section{ CONCLUSION}
\hspace{1.2em}
In this paper, we proposed an improved version 
of the traditional PVO-k algorithm, termed PPVO-k. 
By reconstructing the mapping method, we enabled 
the embedding of encrypted information into the top k 
pixels with the highest and equal grayscale values 
within each image block. This enhancement can 
significantly boost the performance of our algorithm 
in terms of embedding capacity, particularly in smooth 
images. However, the algorithm also has limitations; 
its performance on coarse images in terms of the 
embedding capacity needs improvement. In the future, 
our research will focus on developing improved 
embedding strategies for the PPVO-k algorithm on 
coarse images and designing an adaptive planning 
algorithm to select the optimal mapping strategy

\end{multicols}

\end{document}